\documentstyle[preprint,eqsecnum,aps]{revtex}
\begin{document}
\draft
\preprint{HYUPT-95/1
         \hspace{-28.5mm}\raisebox{2.4ex} {SNUTP 95-019}}
\title{Noncovariant Local Symmetry in Abelian Gauge Theories}
\author{Hyun Seok Yang and Bum-Hoon Lee}
\address{Department of Physics, Hanyang University, Seoul 133-791, Korea}
\maketitle

\begin{abstract}
\hskip 1.0cm We find noncovariant local symmetries in the Abelian
gauge theories. The N\"other charges generating these symmetries are
nilpotent as BRST charges, and they impose new constraints on the
physical states.
\end{abstract}
%\narrowtext
\section{Introduction}
\label{sec:intro}
\hskip .5cm
The theory of gauge fields is based on symmetry principles and the
hypothesis of locality of fields. The principle of local gauge
invariance determines all the forms of the interactions and allows the
geometrical description of the interactions \cite{Utiy56}.
But the quantization of gauge
fields leads to difficulties due to the constraints arising from the
gauge symmetry. These difficulties of the quantization of
constrained systems can be circumvented by the extension of
phase space including the anticommuting ghost variables.
In this approach, the original gauge symmetry is transformed into the
so-called BRST symmetry in the extended phase space \cite{BRST,BFV,Henn85}.
Thus, one expects this BRST symmetry must have the same role as the original
gauge symmetry. In other words, the BRST symmetry must
determine all the forms
of the interactions and the algebraic and topological properties
of the fields in
the quantum theory \cite{Baul85}. This extension of phase (or configuration)
space may imply a larger class of symmetry in an extended phase
(or configuration) space.
Indeed, recent works \cite{Lave93,Tang94} reported the existence
of new and larger symmetries in QED.

In this paper, we find the noncovariant local symmetries in Abelian gauge
theories, these symmetries are similar to the symmetry of Ref. 6
discovered in QED, but ours are local. Our symmetries are not
included in the class of the symmetry of Ref. 7
since they are non-covariant. In addition, we find that
the N\"other charges generating these symmetries are nilpotent as
BRST charges and that they impose strong constraints on state space.

\section{New BRST-like symmetry in Abelian gauge theories}

\newcommand{\be}{\begin{equation}}
\newcommand{\ee}{\end{equation}}
\newcommand{\bea}{\begin{eqnarray}}
\newcommand{\eea}{\end{eqnarray}}
\newcommand{\ba}{\begin{array}}
\newcommand{\ea}{\end{array}}

\hskip .5cm
Consider the BRST and anti-BRST invariant effective QED Lagrangian.
(Our BRST treatments are parallel with those of Baulieu's paper
\cite{Baul85}.)
\bea
{\cal L}_{eff} = &-&\frac{1}{4}F_{\mu\nu}F^{\mu\nu}+ {\bar\psi}
(i\gamma^{\mu}D_{\mu}-m)\psi-\frac{1}{2}
{\bar s}s(A_{\mu}^2+\alpha{\bar c}c) \nonumber\\
= &-&\frac{1}{4}F_{\mu\nu}F^{\mu\nu}+ {\bar\psi}
(i\gamma^{\mu}D_{\mu}-m)\psi +
A_{\mu}\partial^{\mu}b+\frac{\alpha}{2}b^2
-\partial_{\mu}{\bar c}\partial^{\mu}c\label{qedl}
\eea
where $D_{\mu}= \partial_{\mu}+ieA_{\mu}$ is the covariant derivative with
the metric $g_{\mu\nu}=(1,-1,-1,-1).$
This effective Lagrangian has rigid symmetries under the following
BRST and anti-BRST transformations:
\bea
\ba{ll}
sA_{\mu}=\partial_{\mu}c,\;\;
& {\bar s}A_{\mu}=\partial_{\mu}{\bar c}, \\
sc=0,\;\; &  {\bar s}{\bar c}=0,           \\
s{\bar c}=b,\;\; &  {\bar s}c=-b, \\ \label{qedt}
s\psi=-iec\psi, \;\; &  {\bar s}\psi=-ie{\bar c}\psi,  \\
sb=0, \;\; &  {\bar s}b=0.
\ea
\eea
We have introduced an auxiliary field $b$
to achieve off-shell nilpotency of the
BRST and the anti-BRST transformations.
Under the transformations in Eq. (\ref{qedt}),
the original gauge invariant classical Lagrangian remains
invariant while the variation from the gauge-fixing term in the effective
Lagrangian in Eq. (\ref{qedl}), i.e., $A_{\mu}\partial^{\mu}b+
\frac{\alpha}{2}b^2 \approx -\frac{1}{2\alpha}(\partial_{\mu}A^{\mu})^2$,
is canceled by the variation from the ghost term.

The nilpotent N\"other charges generated by the BRST
and the anti-BRST symmetries read as
\bea
Q &=& \int d^3x \{-(\partial_i F^{io}-\rho)c-b\dot{c}\},\label{qedq}\\
{\bar Q} &=& \int d^{3}x \{-(\partial_i F^{io}-\rho){\bar c}-
b\dot{\bar{c}}\}\label{qedaq}
\eea
where $\rho$ is a charge density defind by
\be
\rho=e {\bar \psi}\gamma_0 \psi.\label{qedch}
\ee
The transformations in Eq. (\ref{qedt}) now can be defined as follows:
\[ s{\cal F}(x)=i[Q,{\cal F}(x)\}, \;\; {\bar s}{\cal F}(x)
=i[{\bar Q},{\cal F}(x)\}\]
where the symbol $[\;,\;\}$ is the graded commutator.
In the language of quantum field theory, the BRST charge $Q$ and
the anti-BRST ${\bar Q}$ are the generators of the quantum
gauge transformation.

In order to recover the probabilistic
interpretation of the quantum theory,
we must project out all the physical states in positive
definite Hilbert space. We achieve this goal by asking for BRST
invariance on an extended state space \cite{Naka90}:
\be
Q|phys>=0.\label{phys}
\ee

Now let us consider the BRST-like operators obtained by canonical
transformation interchanging the roles of the ghost $c$ and
the antighost ${\bar c}$ in Eqs. (\ref{qedq}) and (\ref{qedaq})
along the following combinations:
\bea
&& {\bar Q}_c = \int d^{3}x \{(\partial_i F^{io}-\rho)\dot{{\bar c}}
+b\nabla^2 {\bar c}\},\label{qedcbrch}\\
&& {Q}_c = \int d^{3}x \{(\partial_i F^{io}-\rho)\dot{c}
+b\nabla^2 c\}.\label{qedcbr}
\eea
These operators are also nilpotent, i.e.,
${\bar Q}^{2}_{c} = {Q}^{2}_{c}=0$. In Ref. 9, we presented
the geometrical meaning of the operator ${\bar Q}_{c}$, which can be
interpreted as an {\it adjoint} operator of the BRST operator $Q$ based on
Lie algebra cohomology.
We find that $Q({\bar Q})$ and ${\bar Q}_c(Q_c)$ satisfy the
supersymmetry-like algebra that closes into an operator $\Delta$,
\bea
\ba{lll}
\{Q,{\bar Q}_c\} = i\Delta,\;\; & [\Delta, Q]=0,\;\;
&[\Delta, {\bar Q}_c]=0,\\
\{{\bar Q},Q_c\} = -i\Delta,\;\; &[\Delta, {\bar Q}]=0,\;\;
&[\Delta, Q_c]=0,\\
\{Q,{\bar Q}\} =0,\;\; & \{Q, Q_c\}=0,\;\; &\{{\bar Q}, {\bar Q}_c\}=0,
\label{bra}
\ea
\eea
where the operator $\Delta$ can be expressed in terms of the
constraint functions as follows:
\be
\Delta = \int d^{3}x \{(\partial_i F^{io}-\rho)^2+(\nabla b)^2\}.
\label{qedlap}
\ee

The transformation generated by the operator ${\bar Q}_c$ is defined as
${\bar s}_c {\cal F}(x)=i[{\bar Q}_c,{\cal F}(x)\}$,
and the explicit transformations are
\begin{eqnarray}
 \ba{ll}
 {\bar s}_cA_{0}=-\nabla^2 {\bar c}, \;\;
 & {\bar s}_cA_i=-\partial_0\partial_{i}{\bar c},\\
 {\bar s}_c c=(\partial_i F^{io}-\rho),\;\;
 & {\bar s}_c{\bar c}=0, \\ \label{qedct}
 {\bar s}_c\psi=ie\dot{\bar c}\psi, \;\;
 & {\bar s}_c{\bar\psi}=-ie\dot{\bar c}{\bar\psi},\\
 {\bar s}_cb=0.
\ea
\eea
Note that the nilpotency of the transformation acting on
the ghost field $c$ holds only on-shell.

Remarkably, we can easily show that this noncovariant local transformation
is a symmetry of the Lagrangian in Eq. (\ref{qedl}).
Under the transformation in Eq. (\ref{qedct}),
the Lagrangian in Eq. (\ref{qedl}) changes by a total divergence
\[ {\bar s}_c{\cal L}_{eff}=\partial_{\mu}\Lambda^{\mu}\]
where
\[\Lambda^{0}=(\partial_i F^{io}-\rho)\dot{{\bar c}}-b\nabla^2 {\bar c},\]
\[\Lambda^{i}=(\partial_j F^{jo}-\rho)\partial^i{{\bar c}}
-b\partial^i\dot{{\bar c}}-F^{i0}\Box{\bar c}.\]
It can be shown that this symmetry generates the same kinds of
noncovariant Ward identities as those of Ref. 6 except that
our identities are multiplied by ${\bf q}^2$.
Notice the symmetry in Eq. (\ref{qedct}) still holds true off-shell
regardless of the gauge parameter $\alpha$.

The symmetry generated by the transformation in Eq. (\ref{qedct})
is realized as
the following combination: the gauge-fixing term in the effective
Lagrangian in Eq. (\ref{qedl}), i.e., $A_{\mu}\partial^{\mu}b+
\frac{\alpha}{2}b^2 \approx -\frac{1}{2\alpha}(\partial_{\mu}A^{\mu})^2$,
remains invariant under the transformation and the variation from
the ghost term is canceled up to the total derivative by the variation
from the original gauge-invariant classical Lagrangian.
We would like to draw the reader's attention to different combinations
realizing the symmetries between the (anti-)BRST symmetry and
the new (anti-)BRST-like symmetry being discussed above.
Compared to the BRST symmetry, it is a very interesting property.
In a later paper \cite{Yang2},
we will discuss its physical meaning and the Ward identity
of this symmetry based on the path integral approach,
especially focusing on differences from the BRST symmetry.

With the interchange ${\bar c} \rightarrow c$,
we can also obtain the antiform of the symmetry in Eq. (\ref{qedct})
generated by
the charge $Q_c$. In the following, we want to focus only on the symmetry
generated by the operator ${\bar Q}_c$ because the antiform of the
symmetry generated by the charge $Q_c$ can be obtained by the trivial
substitution ${\bar c} \rightarrow c$.

Since the effective QED Lagrangian, Eq. (\ref{qedl}), is invariant under the
transformation in Eq. (\ref{qedct}),
the physical state $|\Psi>$ must also satisfy
the following condition \cite{Lave93,Tang94}:
\be
{\bar Q}_c|\Psi>=0.\label{cphys}
\ee
If we have more than one supplementary condition, such as Eq. (\ref{phys})
and Eq. (\ref{cphys}), we can deduce further supplementary conditions
from them by taking (anti-)commutators of the operators in them, i.e.,
\be
\{Q,{\bar Q}_c\}|\Psi>= i\Delta|\Psi>=0 \;\Leftrightarrow
\int d^{3}x \{(\partial_i F^{io}-\rho)^2+(\nabla b)^2\}|\Psi>=0.
\label{harm}
\ee
There is no more supplementary condition due to Eq. (\ref{bra}).
Since the operator $\Delta$ consists of non-negative operators and
the condition in Eq. (\ref{harm}) must be satisfied everywhere,
the supplementary condition in Eq. (\ref{harm}) can be rewritten
as two conditions:
\be
(\partial_i F^{io}-\rho)|\Psi>=0 \;\;\mbox{and}
\;\;b|\Psi>=0. \label{qedharms}
\ee
Evidently these constraints on the physical state
automatically satisfy the previous constraints in Eqs. (\ref{phys})
and (\ref{cphys}) while the former conditions can contain additional
states which are annihilated by the ghost operators \cite{Neme88}.
These two conditions in Eq. (\ref{qedharms}) are the same ones as Dirac's
supplementary conditions \cite{Dirac58} or the Gupta-Bleuler
condition in the gauge-fixing function $b$ \cite{Naka90}.

The same kinds of symmetries as in QED also exist
in the BRST Landau-Ginzburg Lagrangian and the BRST Chern-Simons
Lagrangian. The argument about the Landau-Ginzburg theory is exactly equal
to that of QED, except for a trivial modification of the matter part,
so we do not repeat it here.

Now we present the noncovariant local symmetries in the BRST Chern-Simons
Lagrangian in (2+1)-dimension:
\be
{\cal L}_{eff} = \frac{\kappa}{4}\epsilon^{\mu\nu\lambda}A_{\mu}F_{\nu\lambda}
+ |D_{\mu}\phi|^2+A_{\mu}\partial^{\mu}b+\frac{\alpha}{2}b^2
-\partial_{\mu}{\bar c}\partial^{\mu}c.\label{csl}
\ee
This effective Lagrangian is invariant under the BRST
and the anti-BRST transformations, which is the same form
as Eq. (\ref{qedt}) for QED.
The nilpotent N\"other charges generated by the BRST
and the anti-BRST symmetries read as
\bea
Q &=& \int d^2x \{-(\frac{\kappa}{2}\epsilon^{ij}F_{ij}-\rho)c
-b\dot{c}\},\label{csq}\\
{\bar Q} &=& \int d^{2}x \{-(\frac{\kappa}{2}\epsilon^{ij}F_{ij}-\rho)
{\bar c}-b\dot{\bar{c}}\}\label{csaq}.
\eea
Then, the nilpotent BRST-like charges ${\bar Q}_c$ and $Q_c$
in the Chern-Simons theory are given by
\bea
&& {\bar Q}_c = \int d^{2}x \{\frac{\kappa}{2}\epsilon^{ij}F_{ij}
-\rho)\dot{{\bar c}}+b\nabla^2 {\bar c}\},\label{cscq}\\
&&Q_c = \int d^{2}x \{\frac{\kappa}{2}\epsilon^{ij}F_{ij}
-\rho)\dot{c}+b\nabla^2 c\},\label{cscqc}
\eea
and the transformations generated by the operator ${\bar Q}_c$ are
\begin{eqnarray}
\ba{ll}
{\bar s}_cA_{0}=-\nabla^2 {\bar c}, \;\;
 & {\bar s}_cA_i=-\partial_0\partial_{i}{\bar c},\\
{\bar s}_c c=(\frac{\kappa}{2}\epsilon^{ij}F_{ij}-\rho), \;\;
 & {\bar s}_c{\bar c}=0, \\ \label{csct}
{\bar s}_c\phi=ie\dot{\bar c}\phi, \;\;
 & {\bar s}_c\phi^{\ast}=-ie\dot{\bar c}\phi^{\ast},\\
{\bar s}_cb=0.
\ea
\eea
The Lagrangian in Eq. (\ref{csl}) changes by a total divergence under the
transformations in Eq. (\ref{csct}):
\[ {\bar s}_c{\cal L}_{eff}=\partial_{\mu}\Lambda^{\mu}\]
where
\[\Lambda^{0}=(\frac{\kappa}{2}\epsilon^{ij}F_{ij}-\rho)\dot{{\bar c}}
-b\nabla^2 {\bar c}-\frac{\kappa}{2}\epsilon^{ij}
A_i\partial_j\dot{{\bar c}},\]
\[\Lambda^{i}=(\frac{\kappa}{2}\epsilon^{jk}F_{jk}-\rho)\partial^i{\bar c}
-b\partial^i\dot{{\bar c}}-\kappa\epsilon^{ij}A_j\Box{\bar c}
+\frac{\kappa}{2}A_0\epsilon^{ij}\partial_j\dot{{\bar c}}.\]
The symmetry in Eq. (\ref{csct}) still holds true off-shell, regardless of
the gauge parameter $\alpha$.
Of course, with the interchange ${\bar c} \rightarrow c$,
we can also obtain the antiform of the symmetry in Eq. (\ref{csct})
generated by the charge $Q_c$.

Since the Chern-Simons Lagrangian, Eq. (\ref{csl}), is invariant
under the BRST transformation and the transformations in Eq. (\ref{csct}),
the physical state $|\Psi>$ of the Chern-Simons theory must satisfy
the conditions
\be
Q|\Psi>={\bar Q}_c|\Psi>=0.\label{csphys}
\ee
These conditions induce an additional constraint on the physical state
$|\Psi>$:
\be
\Delta|\Psi>=0 \; \Leftrightarrow
\int d^{2}x \{(\frac{\kappa}{2}\epsilon^{ij}F_{ij}-\rho)^2
+(\nabla b)^2\}|\Psi>=0.
\label{csharm}
\ee
Consequently, the physical state $|\Psi>$ of the Chern-Simons theory
must satisfy the following two conditions, as it does in the case of QED:
\be
(\frac{\kappa}{2}\epsilon^{ij}F_{ij}-\rho)|\Psi>=0
\;\; \mbox{and}\;\; b|\Psi>=0.\label{cshs}
\ee
These conditions are the familiar ones with the anyon problem in
the Chern-Simons matter systems \cite{Fort92,Yang3}.
In Ref. 14, we will discuss the fact that the supplementary
conditions in Eq. (\ref{cshs}) have an important role in
the construction of a physical anyon operator in a covariant gauge.

\section{Some remarks on the BRST-like symmetry}

\hskip .5cm
In Sec. II, we presented a noncovariant local BRST-like symmetry
in Abelian gauge theories. This symmetry is still compatible with
the gauge-fixing condition, i.e., ${\bar s}_c(\partial_{\mu}A^{\mu})=0$,
and there is no reason to abandon locality, unlike the claim in Ref. 6.

Notice the transformations in Eqs. (\ref{qedct})
and (\ref{csct}) can be deformed as follows:
\begin{eqnarray}
 \ba{ll}
 {\bar s}_cA_{\mu}=-\partial_0\partial_{\mu}{\bar c},\\
 {\bar s}_c c=-\dot{b},\;\;& {\bar s}_c{\bar c}=0,\\
 {\bar s}_cb=0,\\
 {\bar s}_c\psi(\phi)=ie\dot{\bar c}\psi(\phi), \;\;
 & {\bar s}_c{\bar\psi}(\phi^*)=
   -ie\dot{\bar c}{\bar\psi}(\phi^*).\label{dct}\\
\ea
\eea
One can easily verify, under the above transformations,
that the effective actions in Eqs. (\ref{qedl}) and (\ref{csl})
also change by a total divergence, like the previous ones.
Thus, the above transformations represent another local symmetry
existing in the Abelian gauge theories, which is equivalent to the
symmetries in Sec. II only on-shell. This symmetry still preserves the
locality and corresponds to the localized version of Eq. (3)
of Ref. 7.

Based on the existence of the local symmetry above,
one may also expect a nonlocal symmetry corresponding to the
local one as long as we demand good boundary conditions on the
fields \cite{Lave93} (see also Ref. 7, especially Eq. (4)).
In fact, the nonlocal symmetries in Refs. 7 and 8 are examples
showing the existence under the special
requirement with respect to the operators generating the
nonlocality.

In a recent work \cite{Yang4}, we have shown that there exists a
straightforward way to isolate the physical Hilbert space
with a positive-definite norm through the BRST cohomology.
In that paper, we introduced the ``co-BRST'' operator under
a positive-definite inner product and obtained the Hodge
decomposition theorem in the state space of BRST quantization.
The co-BRST operator in Ref. 15 is quite different from
the operator ${\bar Q}_c$ discussed in this paper.
The BRST-like charges, Eqs. (\ref{qedcbrch}) and (\ref{qedcbr}), can be used
to refine the characterization of the
physical states as in Eq. (\ref{cphys}) and Eq. (\ref{qedharms}), but they
cannot be applied to the problem to directly isolate the physical state
with a positive-definite norm in the sense of Ref. 15.

For the case of free QED, we have written down the explicit form of
the co-BRST operator using the mode expansion of the field operators and
have shown it to be a conserved operator.
If the BRST Hamiltonian in the system under
consideration is Hermitian, the co-BRST operator introduced
in our paper is, in general, a conserved charge
that commutes with the BRST Hamiltonian. Thus, there may exist
a symmetry transformation generated by the co-BRST operator.
However, it seems it is very difficult to express the co-BRST
transformation in an explicit form in configuration space, even
in free theory and it will be different from the symmetry discussed in
this paper and in Refs. 6 and 7.
For that reason, it will be a very interesting problem
to invesigate a larger class
of symmetry including the co-BRST symmetry in gauge theories.

\section*{ACKNOWLEDGEMENTS}
This work was supported by the Korean Science and Engeneering Foundation
(94-1400-04-01-3 and Centor for Theoretical Physics) and
by the Korean Ministry of Education (BSRI-94-2441).


\newpage
\begin{thebibliography}{99}

\bibitem{Utiy56}R. Utiyama, Phys. Rev. {\bf 101}, 1597 (1956).
\bibitem{BRST} C. Becchi, A. Rouet, and R. Stora, Comm. Math. Phys.
{\bf 42}, 127 (1975); Ann. Phys. (N.Y.) {\bf 98}, 287 (1976);
I. V. Tyutin, Lebedev preprint, FIAN  No. {\bf 39} (1975), unpublished.
\bibitem{BFV} E. S. Fradkin and G. A. Vilkovsky, Phys. Lett. B{\bf 55},
224 (1975); I. A. Batalin and G. A. Vilkovsky, Phys. Lett. B{\bf 69},
309 (1977).
\bibitem{Henn85}M. Henneaux, Phys. Rep. {\bf 126}, 1 (1985).
\bibitem{Baul85}L. Baulieu, Phys. Rep. {\bf 129}, 1 (1985).
\bibitem{Lave93}M. Lavelle and D. McMullan, Phys. Rev. Lett.
{\bf 71}, 3758 (1993).
\bibitem{Tang94}Z. Tang and D. Finkelstein, Phys. Rev. Lett.
{\bf 73}, 3055 (1994).
\bibitem{Naka90}N. Nakanishi and I. Ojima, {\it Covariant Operator Formalism
of Gauge Theories and Quantum Gravity} (World Scientific, Singapore, 1990).
\bibitem{Yang1}H. S. Yang and B.-H. Lee, hep-th/9503204.
\bibitem{Yang2}H. S. Yang and B.-H. Lee, in preparation.
\bibitem{Neme88}D. Nemeschansky, C. Preitschopf, and M. Weinstein,
Ann. Phys. (N.Y.) {\bf 183}, 226 (1988).
\bibitem{Dirac58}P. A. M. Dirac, {\it The Principles of Quantum Mechanics},
 4th ed. (Oxford Univ. Press, London, 1958).
\bibitem{Fort92}S. Forte, Rev. Mod. Phys. {\bf 64}, 193 (1992) and
references therein.
\bibitem{Yang3}H. S. Yang and B.-H. Lee, in preparation.
\bibitem{Yang4}H. S. Yang and B.-H. Lee, J. Korean Phys. Soc.
{\bf 28}, 138 (1995).

\end{thebibliography}
\end{document}